\documentclass[12pt,letterpaper]{article}
\pdfoutput=1

\usepackage{jheppub}
\usepackage{amsmath}
\usepackage{amssymb}
\usepackage{xcolor}
\usepackage{graphicx}
\usepackage{subfig}
\usepackage{bbold}
\usepackage{tabularx}
\usepackage[makeroom]{cancel}
\usepackage{multirow}
\usepackage{slashed}
\usepackage{hyperref}
\usepackage{mathtools}
\usepackage{dirtytalk}
\usepackage{tabularx}
\usepackage{rotating}
\usepackage{nicematrix}
\usepackage{caption}
\usepackage{comment}
\usepackage{array}

\title{Compact space catalysis of false vacuum decay and Schwinger effect}

\author[a,b]{Saquib Hassan}

\author[a]{and John March-Russell}

\emailAdd{saquib.hassan@physics.ox.ac.uk}

\emailAdd{john.march-russell@physics.ox.ac.uk}

\affiliation[a]{Rudolf Peierls Centre for Theoretical Physics, University of Oxford, Oxford, OX1 3PU, United Kingdom}
\affiliation[b]{Christ Church, University of Oxford, Oxford, OX1 1DP, United Kingdom}

\date{\today}

\newcommand{\pd}{\partial}

\definecolor{darkgreen}{RGB}{50,150,0}

\definecolor{darkred}{RGB}{150,0,50}

\definecolor{darkblue}{RGB}{0,50,150}

\abstract{
We study zero-temperature false vacuum decay in $D$ compact spatial dimensions and show that for volumes below a critical value a new bounce solution, different from Coleman's celebrated $O(D)$ bubble, mediates the decay process, and typically leads to an exponentially enhanced decay rate. The bounce, when analytically continued to Lorentzian signature, nucleates a homogeneous field configuration for spatial volumes below a critical value, and quasi-homogeneous configurations for slightly larger volumes, and is not of the form of a thin or thick-walled bubble embedded in a false vacuum background. We explicitly show that the new bounce has the necessary features associated with false vacuum decay, following from its eigenvalue spectrum of fluctuations. The cross-over from homogeneous to quasi-homogeneous solutions as the spatial volume is increased is discussed, as is a real-time interpretation of the bounce. We apply this bounce to the study of a scalar field model, as well as a close cousin of the Schwinger effect that applies to $(1+1)d$ axion electrodynamics in compact space.
}

\begin{document}

\maketitle

\section{Introduction}
\label{sec:introduction}

Decays of metastable ``vacuum" states in finite space tend to have peculiar properties which are unexpected from usual calculations in infinite space. Whilst many decay processes and rates in quantum field theory tend to be altered by finite size, background geometry, or temperature effects such as \cite{brown2018schwinger,Hawking:1981fz,Mogliacci:2018oea, Coleman:1980aw,Linde:1981zj, Garriga:1994ut, PhysRevD.46.5321,Shkerin:2021zbf,Geller:2026pup}, in this paper we are particularly interested in how zero-temperature false vacuum decay rates are modified by the effects of finite spatial volume. The Schwinger effect \cite{Heisenberg:1936nmg,Schwinger:1951nm} in finite space \cite{brown2018schwinger} is of special interest. Indeed, when the spatial volume becomes smaller than a certain critical scale, decay rates often tend to be exponentially enhanced. This point is especially intriguing, since one may view the Schwinger effect as a type of false vacuum decay process.

While in infinite space, the decay rate per volume of an electric field of magnitude $E$ through particle-antiparticle pairs, each of mass $m$ and charge $\pm q$, is given by 
\begin{equation}\label{eq:SchwingerRate}
    \Gamma/V \sim \exp{\Bigg(-\frac{\pi m^2}{qE}\Bigg)},
\end{equation}
the recent work of \cite{brown2018schwinger} shows instead that when space is compact (in this case in $(1+1)d$ with space compactified on a circle of perimeter $L$), the decay rate becomes
\begin{equation}
    \Gamma \sim \exp{\Bigg(-2mL\Big(1-\frac{L^2}{24 x_\star^2}+\cdots\Big)\Bigg)},
\end{equation}
where $x_\star =m/qE$, is the radius of the critical bubble in infinite space; in this case, half the separation between the charged pair at nucleation. One assumes throughout that $L\lesssim x_\star$. We review these results in appendix \ref{sec:ScwhingerCompactSpace}. It follows that when space is compactified on extremely small scales, the rate becomes approximately\footnote{Semiclassical validity still requires $L\gtrsim m^{-1}$.}:
\begin{equation}\label{eq:SchwingerRateCompact2}
    \Gamma\sim \exp{\Big(-2mL\Big)}.
\end{equation}

This result is notably different from the naive expectation that as the physical space becomes smaller, the vacuum becomes stable because a bubble mediating the decay no longer fits. Instead, we find that the decay rate grows\footnote{See also \cite{Qiu:2020gbl}.}, and is independent of $x_\star$. Importantly, the action now has an explicit linear scaling in volume (length in this case). There is an important qualitative lesson here. Whereas on the infinite line, a larger electric field decays via a charged pair with the electric field in the interior region enclosed by the pair weaker than the field in the exterior, for a small circle instead, \cite{brown2018schwinger} argues that a stronger electric field decays into a weaker one across the entire space directly, producing only a pair of photons via virtual charged pair annihilation, since no real charged pair is formed. Intuitively, the larger electric field decays \textit{homogeneously} into a weaker one, which also justifies the volume scaling of the action in eq. \eqref{eq:SchwingerRateCompact2}. This observation will be important when we generalize false vacuum decay in compact space to quantum field theory.

This consideration leads us to explore how the false vacuum decay in finite spatial volumes could be different from Coleman's bubble nucleation story \cite{Coleman:1977py,Callan:1977pt}, and in doing so we attempt to place results such as eq. \eqref{eq:SchwingerRateCompact2} on a more unified footing. We find that even when the linear size of all spatial directions is smaller than the size of Coleman's critical bubble, and naively the energetics of true-vacuum bubble solutions appears to forbid bubble nucleation, the false vacuum is nonetheless able to decay. This process occurs through a different homogeneous or quasi-homogeneous bounce solution with no region of the compact space left in the false vacuum basin.  This new homogeneous bounce leads to a qualitatively different and exponentially enhanced decay rate.  


The structure of the paper is as follows. In section \ref{sec:compact} we will introduce the homogeneous bounce solution that mediates false vacuum decay in small compact spaces (we also explore how compact size and spatial curvature effects modify the usual decay rate when space is larger than the critical Coleman bubble diameter, at least in one particular example, in appendix \ref{sec:appB}). We find a bounce configuration, extended in space, satisfying key requirements to mediate vacuum decay, such as time reversal symmetry at the bounce center, a single negative eigenvalue for fluctuations, etc. In section \ref{sec:eigen} we solve an explicit example and compute the eigenvalue spectrum of quadratic fluctuations about the homogeneous bounce, and discuss how as the spatial size increases above a critical value new collective coordinate zero modes arise, as well as a mode indicating a bifurcation to a preferred quasi-homogeneous bounce solution.  We also discuss a real time interpretation of our homogeneous bounce. In section \ref{sec:appA} we apply our new bounce to a study a theory with charged domain walls - akin to charged particle pairs in QED - in the context of axion electrodynamics in $(1+1)d$, obtaining a Schwinger rate for the theory in compact space, demonstrating striking similarities with \cite{brown2018schwinger}.

\section{Vacuum decay in compact space through the homogeneous bounce}
\label{sec:compact}

\begin{figure}[h]
\centering\includegraphics[width=0.8\textwidth]{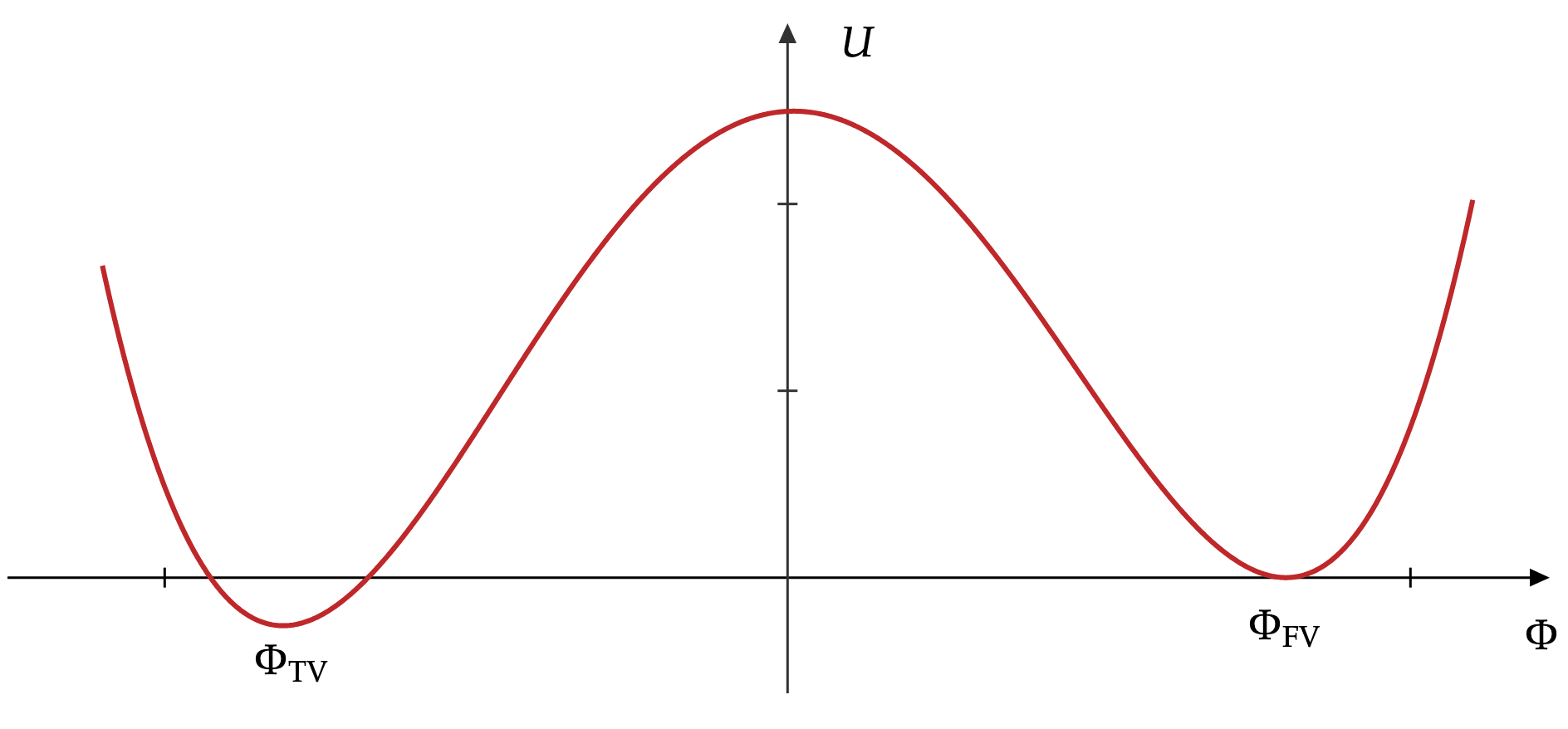}
\caption{The potential $U(\Phi)$ for a scalar field $\Phi$ showing a false vacuum field value $\Phi_{\mathrm{FV}}$ for which the potential vanishes and is a local minimum, and a true vacuum corresponding to a global minimum at $\Phi_{\mathrm{TV}}$.}
\label{fig:potential}
\end{figure}

First, let us quickly outline some of the key features of false vacuum decay (in infinite space and time) that are necessary for the remainder of this paper\footnote{Our presentation here is terse. The reader may consult \cite{Coleman:1985rnk,Coleman:1977py,Callan:1977pt} for details.}. We start in the distant past with the expectation value of a field $\Phi (t,\mathbf{x})$ in the metastable minimum of a potential $U(\Phi)$, corresponding to a constant false vacuum value $\Phi_{\mathrm{FV}}$. It is convenient to set $U(\Phi_{\mathrm{FV}})=0$. There is also a true vacuum field value, $\Phi_{\mathrm{TV}}$, corresponding to a lower minimum with $U(\Phi_{\mathrm{TV}})<0$. For instance, take the action in $D$ dimensional spacetime:
\begin{equation}\label{eq:scalaraction}
    I=\int d^D x\ \Big(-\frac{1}{2}\partial_\mu \Phi \partial^\mu \Phi -U(\Phi)\Big),
\end{equation}
with
\begin{equation}
    U(\Phi)=U_{+}(\Phi)+\frac{\epsilon}{2a}(\Phi-a), \quad U_{+}(\Phi)=\frac{\lambda}{8}\Big(\Phi^2 -\frac{\mu^2}{\lambda}\Big)^2,
\end{equation}
and $\mu^2/\lambda=a^2$. The precise values of the parameters are not particularly important for the present discussion, but note that $\Phi_{\mathrm{FV}}=a$ here. See figure \ref{fig:potential}.

In Euclidean signature, we sum over all paths, taking the initial false vacuum field configuration back to itself at late times. A path, or field configuration, that involves only one detour from the false vacuum value before returning to itself is called a single bounce, characterized by a \textit{bounce solution} whose corresponding action is $I_{\mathrm{B}}$. Assuming such bounces are widely separated in Euclidean time and space (cf, the \textit{dilute instanton gas} aproximation), one finds that the sum over all paths exponentiates. Moreover, one then discovers that the vacuum persistence amplitude, the amplitude for the vacuum to evolve to itself at late times, decays\footnote{The presence of one negative eigenvalue in the eigenvalue spectrum of the quadratic fluctuation operator about the action saddle point is important in identifying the decay. See \cite{Coleman:1985rnk} for details.} with decay rate per volume given by $\Gamma/V\sim \exp{(-I_{\mathrm{B}})}$. 

In real time, a spherical bubble of critical radius spontaneously nucleates with its wall separating the false vacuum exterior from the true vacuum interior\footnote{Here we implicitly take a thin wall bubble. For a thick wall, the bubble interior is initially at some intermediate value of the field $\Phi$ rather than the true vacuum value. Although usually, the bubble interior contains lower energy at nucleation, this need not always be so: one particular exception exists \cite{Hassan:2024nbl}.}, and this wall then accelerates outwards, asymptotically approaching the speed of light, converting a universe of false vacuum to one of true vacuum. In practice, there will be many such spontaneous nucleations, and multiple bubble walls may collide with one another.

We also note a crucial requirement here: the bounce action $I_{\mathrm{B}}$ must be finite. We can see that this requirement implies that the bounce solution has to be localized in both space and Euclidean time\footnote{The paper \cite{Coleman:1977th} showed that such bounce solutions are $O(D)$ symmetric in flat infinite $D$ dimensional Euclidean spacetime, i.e. $\Phi(\tau,\mathbf{x})=\Phi(\rho)$, where $\rho^2=\tau^2+|\mathbf{x}|^2$. However, solutions with other symmetry groups exist in the presence of gravity \cite{Masoumi:2012yy}, and more complicated background fields, interactions, and back-reaction effects - such as in \cite{Hassan:2024nbl} (see also \cite{Ho:2021uem}) - may break the $O(D)$ symmetry. That said, merely increasing the number of scalar fields appears \textit{not} to spoil Coleman's ansatz \cite{Blum:2016ipp}.}. Since the potential at $\Phi_{\mathrm{FV}}$ vanishes, and the action contains an integral of the potential over all space and (Euclidean) time, at large distances (both space and time) from the bounce solution, the field $\Phi$ must approach $\Phi_{\mathrm{FV}}$. Otherwise, the bounce action would be infinite due to an infinite domain integral over a non-vanishing finite potential. Now we discuss how this requirement can be relaxed in cases where space is compact.

\subsection{Changes due to finite size}\label{sec:compactchanges}

We would now like to develop a general framework for studying false vacuum decay in compact space. It is instructive to first consider how the usual flat space false vacuum decay rate is modified when space is curved. This problem has been studied in \cite{Abbott:1987xq,PhysRevD.46.5321}. Briefly, the approach in \cite{Abbott:1987xq} is to consider bubble nucleation at a particular point in (curved) space and expand the metric $g_{\mu\nu}(x)$ about the center of the bubble $(x_0)$ in Riemann normal coordinates:
\begin{equation}\label{eq:Riemann}
    g_{\mu\nu}(x_0+\xi)=\eta_{\mu\nu}-\frac{1}{3}R_{\alpha\mu\beta\nu}(x_0)\xi^\alpha \xi^\beta+\cdots
\end{equation}
where terms involving higher powers of the deviation vector $\xi$ are negligible, provided the curvature is not too large at $x_0$. In the end, one finds that the bounce action is modified:
\begin{equation}\label{eq:curvaturemodification}
    \frac{I_\mathrm{B}}{I_\mathrm{B_0}}\approx 1-\frac{1}{12}R(x_0)r_\star^2 +\cdots,
\end{equation}
where $I_\mathrm{B_0}$ is the bounce action in flat space, $I_\mathrm{B}$ is the new bounce action, $r_\star$ is the critical bubble radius in flat space, and $R(x_0)$ is the Ricci scalar at $x_0$. We can see that on spaces of positive curvature, such as a 2-sphere for instance which has a constant Ricci scalar everywhere, the lower bounce action enhances the decay rate. We explore this point further in appendix \ref{sec:appB}, where we study the effect of curvature even when the radius of the 2-sphere is of the same order as the flat space critical bubble.

Now consider the case in which space is so compact that it is smaller than the critical bubble radius. On the one hand, it seems that the decay process will not occur at all because even a single critical bubble will not fit in space, as Coleman's solution cannot be accommodated. On the other hand, we see from eq. \eqref{eq:curvaturemodification} - as well as the results in appendix \ref{sec:appB} - that as the spatial extent becomes smaller, the decay rate \textit{increases}\footnote{It is convenient to visualize the special case of a space being a 2-sphere, in addition to a time direction. As the sphere becomes smaller, the critical bubble resembles a polar cap on this sphere. An increasing Ricci scalar due to a smaller sphere leads to a lower bounce action (eq. \eqref{eq:curvaturemodification}) and hence a faster decay rate. See appendix \ref{sec:appB}.}. This result is also in agreement with \cite{brown2018schwinger}. These two lines of thought appear to lead to contradictory outcomes. We will now argue that the latter conclusion is correct.

When space is compact, we can modify the requirements we mentioned at the beginning of this section. Indeed, due to the compactness of space, an integral over a non-vanishing potential over all space will still be finite. As such, it is now possible to drop our insistence that at large spatial distances, the scalar field $\Phi$ must approach $\Phi_{\mathrm{FV}}$. Nevertheless, a computation of the vacuum persistence amplitude involves sums over paths which, starting at $\Phi_{\mathrm{FV}}$ in the distant past, return to $\Phi_{\mathrm{FV}}$ in the distant future. The Euclidean time integral remains formally over an infinite domain. It follows that we still need to find a solution for which $\Phi(\tau,\mathbf{x}) \rightarrow \Phi_{\mathrm{FV}}$ when $\tau \rightarrow \pm \infty$. That is to say, we still seek a non-trivial field profile that is localized in Euclidean time. It is precisely such solutions that we will attempt to find. From this point onward, consider only the case in which spatial size is much smaller than what the critical bubble size would be in infinite space.

\subsection{Decay rates from homogeneous solutions}

Consider the Euclidean action for a scalar field in $D=(d+1)$ dimensions,
\begin{equation}\label{eq:euclideanAction}
    I_\mathrm{E}[\Phi]=\int d\tau \ d^d x \ \mathcal{L}_\mathrm{E}(\Phi(\tau, \mathbf{x})).
\end{equation}
The Euclidean Lagrangian density is:
\begin{equation}\label{eq:euclideanL}
    \mathcal{L}_\mathrm{E}(\Phi(\tau, \mathbf{x}))=\frac{1}{2}(\partial_\tau \Phi)^2+\frac{1}{2}(\nabla_\mathbf{x} \Phi)^2 +U(\Phi).
\end{equation}
Evaluating this action on a bounce solution will give the decay rate. Based upon the discussion in section \ref{sec:compactchanges}, we would like to incorporate a solution valid on compact space, where the spatial sizes are much smaller than what would be a critical bubble size in flat infinite space. Looking at the Schwinger effect in compact space, eq. \eqref{eq:SchwingerRateCompact2} indicates a pattern in which the volume of space factors out in the bounce action, which we see in the exponent. This insight leads us to consider solutions for $\Phi(\tau,\mathbf{x})$ that deviate from Coleman's $O(D)$ ansatz. More precisely, we consider a solution of the form $\Phi(\tau, \mathbf{x})=\Phi(\tau)$, i.e. a homogeneous solution. Then the above action reduces to the following:
\begin{equation}
     I_\mathrm{E}[\Phi]\rightarrow V\int d\tau \ \mathcal{L}_\mathrm{E}(\Phi(\tau)),
\end{equation}
where $V$ is the volume of the compact space, which mimics eq. \eqref{eq:SchwingerRateCompact2}. The strategy now is to find a solution to the equation of motion from
\begin{equation}\label{eq:homogeneousL}
    \mathcal{L}_\mathrm{E}(\Phi(\tau))= \frac{1}{2}\dot\Phi^2 +U(\Phi),
\end{equation}
which satisfies the boundary condition $\Phi(\tau) \rightarrow \Phi_{\mathrm{FV}}$ when $\tau \rightarrow \pm \infty$. This problem reduces to one of finding quantum mechanical bounce solutions; see \cite{Coleman:1985rnk} for details.

\begin{figure}[h]
\centering\includegraphics[width=0.98\textwidth]{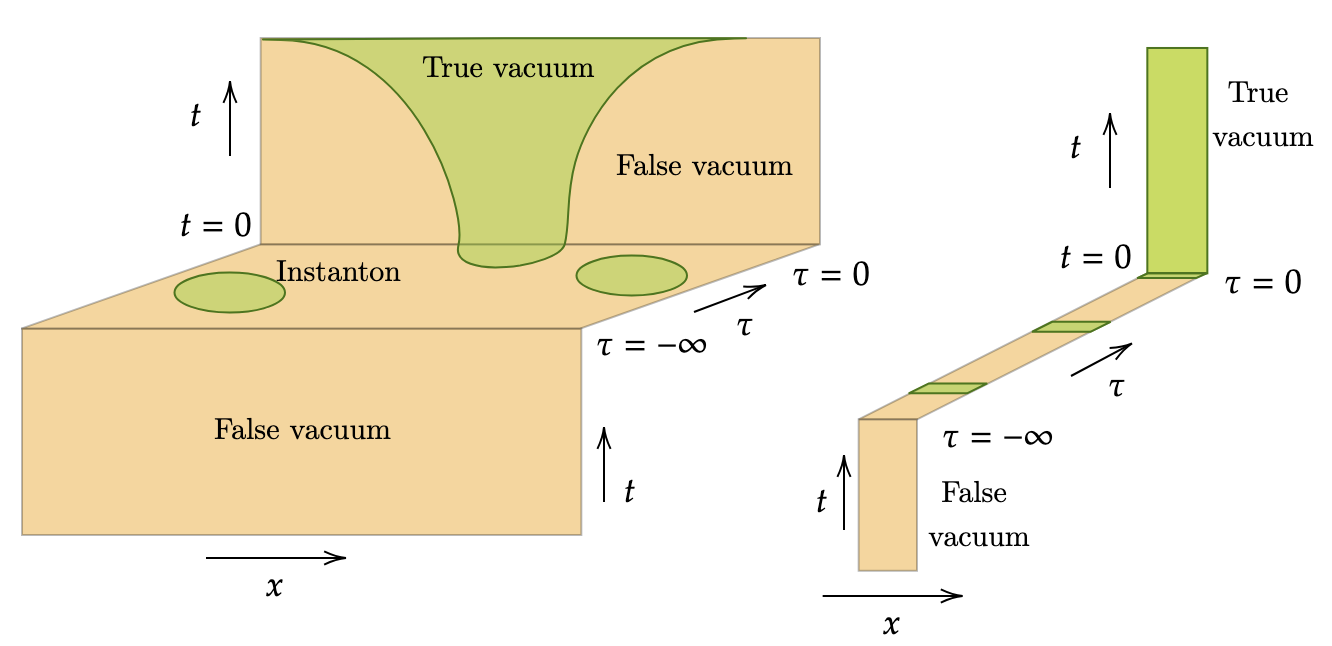}
\caption{Left: The usual $O(D)$ bounce of Coleman. The green bubble represents the true vacuum interior separating the false vacuum exterior shown in beige. Right: The homogeneous bounce in compact space, where the space is compactified on scale smaller than the critical bubble size in infinite space.}
\label{fig:CompactInstanton}
\end{figure}

Before proceeding, we can already visualize what this new solution entails. While in the usual case, a spherical bubble with an interior of true vacuum in (infinite) space spontaneously nucleates before accelerating outwards, now instead the \textit{entire} (finite) space of false vacuum suddenly transitions to true vacuum\footnote{Or rather an intermediate value close to the true vacuum before rolling to the true vacuum classically; we will elaborate upon this subtlety momentarily.}. This time, there is no bubble at all, and therefore no bubble wall that accelerates outwards, in marked contrast with the conventional process. We illustrate this difference in figure \ref{fig:CompactInstanton}. We will now show how this solution is realized.

Varying eq. \eqref{eq:homogeneousL} with respect to $\Phi$ gives the equation of motion
\begin{equation}\label{eq:equationPhiTilde}
   \frac{d^2 \Phi}{d\tau^2}=\frac{dU(\Phi)}{d\Phi}.
\end{equation}

\begin{figure}[h]
\centering\includegraphics[width=0.8\textwidth]{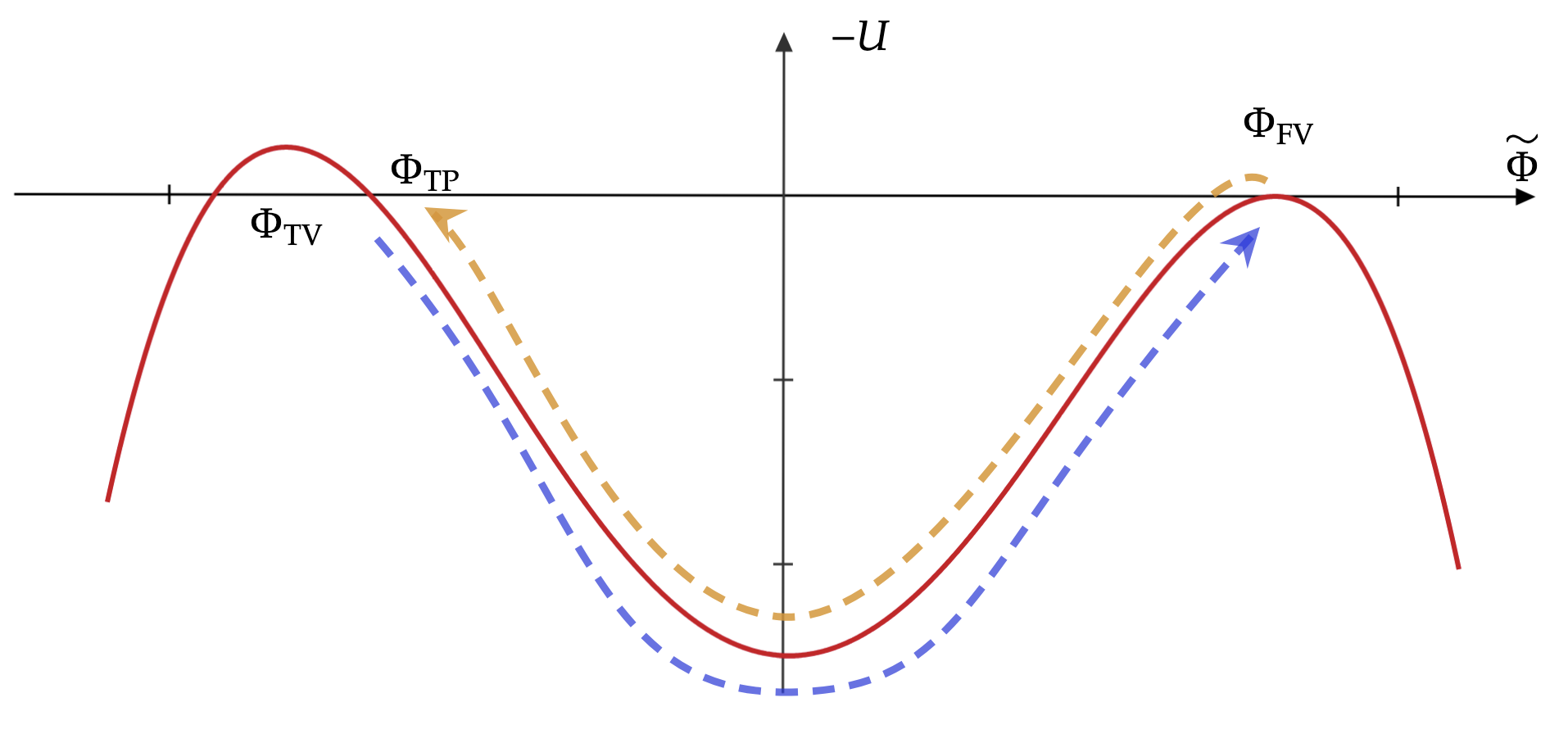}
\caption{The inverted potential (of figure \ref{fig:potential}) corresponding to the bounce solution. The orange dashed line shows motion starting at $\tau=-\infty$ (at $\Phi_{\mathrm{FV}}$) to $\tau=0$ (at $\Phi_{\mathrm{TP}}$, a root of $U(\Phi)$) representing the first half of a single bounce, followed by the return path, shown in the blue dashed line, from $\tau=0$ (at $\Phi_{\mathrm{TP}}$) to $\tau=\infty$ (at $\Phi_{\mathrm{FV}}$) indicating the second half of the same bounce. }
\label{fig:invertedpotential}
\end{figure}

Unsurprisingly, this equation is akin to an $F=ma$ equation for a ball rolling in an inverted potential (figure \ref{fig:invertedpotential}). Unlike the $O(D)$ solution in infinite space, there is no longer any friction/dissipation term and so it is unnecessary to explicitly integrate this equation. Indeed, this system now mimics a purely quantum mechanical bounce in $(0+1)d$. Multiplying eq. \eqref{eq:equationPhiTilde} by $\dot{\Phi}$, we get the following conservation equation that the bounce solution must obey:
\begin{equation}\label{eq:energyConservation}
    \frac{1}{2}\dot{\Phi}^2-U(\Phi)=0,
\end{equation}
where we have used the convention $U(\Phi_{\mathrm{FV}})=0$, as well as the boundary condition above. The bounce solution contains a turning point, $\Phi_{\mathrm{TP}}$, centered at $\tau=0$ at which point $\dot{\Phi}$ vanishes\footnote{Note that this condition is automatically consistent with time reflection symmetry at the center of the bounce.}. But from eq. \eqref{eq:energyConservation}, this field value corresponds to a root of the potential, i.e. the root in between the false vacuum and the true vacuum\footnote{In the usual case of false vacuum decay in infinite space, the friction term in the Euclidean equation of motion renders it such that the turning point at which $\Phi'(\rho)=0$ does not correspond to a root of the potential. So while the true vacuum and the false vacuum remain the same whether we are in compact space or infinite space due to having the same potential, the turning point for the field value will be different.}. See figure \ref{fig:invertedpotential}. Call this bounce solution $\tilde{\Phi}(\tau)$.

Putting eq. \eqref{eq:energyConservation} in the first term on the right hand side of eq. \eqref{eq:homogeneousL}, we immediately obtain the decay rate of the false vacuum:
\begin{equation}\label{eq:compactrate}
        \Gamma\sim \exp{\Bigg( -2V\int_{\Phi_{\mathrm{TP}}}^{\Phi_{\mathrm{FV}}}d\tilde{\Phi}\sqrt{2U(\tilde{\Phi})} \Bigg)},
\end{equation}
where $V$ is the volume of the compact space. The explicit factoring of the volume in the exponent shows that as the volume decreases, the rate increases. Despite appearances, there is no assumption here of a large barrier height. The doubling of the exponent stems from the fact that the above integral captures only half the bounce (figure \ref{fig:invertedpotential}). The resemblance with the WKB approximation is expected if we view the action for a homogeneous field as that of a particle in non-relativistic quantum mechanics, with its coordinate $x$ being $\tilde{\Phi}$.

There is a non-trivial test of eq. \eqref{eq:compactrate} in which we can compute the Schwinger effect in field theory and reproduce what we expect from eq. \eqref{eq:SchwingerRateCompact2}. This calculation is the subject of section \ref{sec:appA}. Finally, recall that the consistency requirements here are that the volume of space be much less than the size of the critical bubble in infinite space, as well as the requirement that the bounce action in the exponent of eq. \eqref{eq:compactrate} be much larger than unity so that we may justify the semiclassical approximation.

\section{On the eigenvalue spectrum near the saddle point}\label{sec:eigen}

There is an important point that we should address: the role of the eigenvalue spectrum of the quadratic fluctuation operator about the saddle point solution. Because ultimately the computation of the decay rate involves a sum over an arbitrary number of bounces, with each bounce estimated by a saddle point approximation and a Gaussian integral, a determinant associated with the fluctuations about the saddle appears in this sum. Vanishing eigenvalues are associated with collective coordinates for a single bounce solution, and importantly, a single negative eigenvalue is responsible for the exponential decay of the vacuum persistence amplitude  \cite{Coleman:1977py,Callan:1977pt,Coleman:1985rnk,Coleman:1977th}. We will now consider the fluctuations about a homogeneous bounce solution.

\subsection{Quadratic fluctuations}

Recall from \cite{Coleman:1985rnk,Coleman:1977py,Callan:1977pt} that the path integral which computes the partition function in the semiclassical approximation, a sum over saddles, involves an integral over over the fields that denote deviations away from the saddle point solution. Explicitly,
\begin{equation}\label{eq:fieldExpansion}
    \Phi(\tau,\mathbf{x})=\tilde{\Phi}(\tau)+\xi(\tau,\mathbf{x}),
\end{equation}
where we conveniently extract the homogeneous solution. The Euclidean action (eq. \eqref{eq:euclideanAction}) becomes:
\begin{equation}\label{eq:actionQuadratic}
    I_\mathrm{E}[\tilde\Phi,\xi]= V\int d\tau \Big( \frac{1}{2}\dot{\tilde{\Phi}}^2 +U(\tilde{\Phi})\Big)+\int d\tau\  d^dx\ \Bigg( \frac{1}{2}\xi\Big(-\partial_\tau^2-\nabla_{\mathbf{x}}^2+U''(\tilde{\Phi})\Big)\xi \Bigg) +\mathcal{O}(\xi^3).
\end{equation}
The absence of a term that is linear in $\xi$ stems from the fact that we are expanding about an extremum of the action. The expansion about a saddle point configuration - corresponding to a \textit{single} bounce - implies that upon Gaussian integration over the bosonic field $\xi$, the partition function becomes approximately
\begin{equation}
    \mathcal{Z}= \int \mathcal{D} \Phi\ \exp{(-I_\mathrm{E}[\Phi])}\approx \int \mathcal{D} \xi\  \exp{(-I_\mathrm{E}[\tilde{\Phi},\xi])} \rightarrow \mathrm{Det}(\mathcal{O})^{-1/2}\exp{(-I_\mathrm{E}[\tilde{\Phi}])},
\end{equation}
ignoring overall numerical factors.
The integration over bosonic fields has produced a square root of a determinant of an operator $\mathcal{O}$, which in our case is:
\begin{equation}\label{eq:quadOperator}
    \mathcal{O}=-\partial_\tau^2-\nabla_{\mathbf{x}}^2+U''(\tilde{\Phi}(\tau)),
\end{equation}
and is evaluated on $\tilde{\Phi}(\tau)$, the homogeneous bounce solution.
By definition, its eigenvalues $\{\lambda\}$ - corresponding to the eigenfunctions $\{\sigma(\tau,x)\}$ -  satisfy $\mathcal{O} \sigma=\lambda \sigma.$

In the usual picture \cite{Coleman:1985rnk}, $\mathcal{O}$ has a vanishing eigenvalue\footnote{Actually, there can be $D$ vanishing eigenvalues corresponding to the translational symmetries in $D$ dimensional Euclidean space for Coleman's $O(D)$ solution. But that is \textit{not} the solution we are studying here. For the homogeneous bounce, the only translational symmetry is Euclidean time translation.}, a negative eigenvalue\footnote{This negative eigenvalue is necessary to infer a decay rate.}, and a tower of positive eigenvalues.
\begin{figure}[h]
\centering\includegraphics[width=0.7\textwidth]{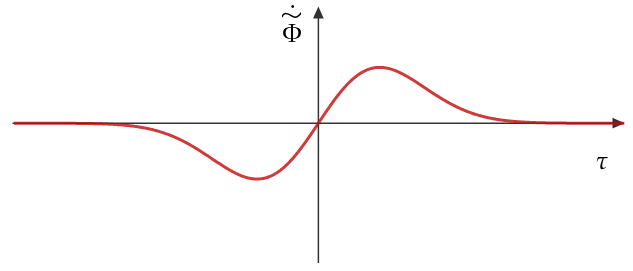}
\caption{The eigenfunction corresponding to zero eigenvalue of the fluctuation operator $\mathcal{O}$. The single node indicates that this eigenfunction has the second lowest eigenvalue. In the distant past $\tau\rightarrow -\infty$, the field starts at $\Phi_{\mathrm{FV}}$ and moves left (negative field derivative) to $\Phi_{\mathrm{TP}}$, arriving at $\tau=0$ at which point the field derivative vanishes at the turning point (this path corresponds to the orange dashed line in figure \ref{fig:invertedpotential}). Then from $\tau=0$ to $\tau=\infty$ the field moves right (positive field derivative), and back to $\Phi_{\mathrm{FV}}$ (blue dashed line of figure \ref{fig:invertedpotential}).}
\label{fig:phidot}
\end{figure}

Let us first discuss the vanishing eigenvalue, and the corresponding eigenfunction. Towards this end, it simplifies matters to first consider the following operator:
\begin{equation}\label{eq:quadOperator2}
   \tilde{\mathcal{O}}=-\partial_\tau^2 + U''(\tilde{\Phi}(\tau)),
\end{equation}
which is the same as $\mathcal{O}$ except for the missing Laplacian. Once we find the eigenvalue spectrum for $\tilde{\mathcal{O}}$, relating it to the spectrum for $\mathcal{O}$ will be a straightforward extension. It is clear that a vanishing eigenvalue of $\tilde{\mathcal{O}}$ exists from time translation symmetry, which is to say that the bounce center may be translated in Euclidean time. Indeed, it is precisely this symmetry that gives the conservation rule of eq. \eqref{eq:energyConservation}, reproduced below:
\begin{equation*}
    \frac{1}{2}\dot{\tilde{\Phi}}^2-U(\tilde{\Phi})=0.
\end{equation*}
Differentiating this relation twice, or differentiating the equation of motion (eq. \eqref{eq:equationPhiTilde}) once, gives
\begin{equation}
   -\frac{d^2\dot{\tilde{\Phi}}}{d\tau^2}+U''(\tilde{\Phi})\dot{\tilde{\Phi}}=0.
\end{equation}
Comparing this relation with eq. \eqref{eq:quadOperator2}, we see that $\sigma\sim \dot{\tilde{\Phi}}$ is an eigenfunction of $\tilde{\mathcal{O}}$ (and of $\mathcal{O}$) with vanishing eigenvalue\footnote{We typically set aside this vanishing eigenvalue as it gives spurious singularities; we can also regulate this effect by carrying out the temporal integral over a finite domain.}. We plot this eigenfunction in figure \ref{fig:phidot}. 

The eigenfunction with vanishing eigenvalue has exactly one node\footnote{It is helpful to have in mind the inverted potential of figure \ref{fig:invertedpotential}.}, and it must therefore be the eigenfunction with the \textit{second} lowest eigenvalue of $\tilde{\mathcal{O}}$. But if the second lowest eigenvalue is zero, then it follows that there is exactly one negative eigenvalue. It is this negative eigenvalue that represents the unstable direction about the saddle point inducing vacuum decay. Coleman argued \cite{Coleman:1977py,Callan:1977pt} that the negative eigenvalue corresponds to an unstable direction associated with dilating the bubble solution when the configuration is that of a thin wall. We will attempt to make an analogous argument below for compact space, and see when a negative eigenvalue gives a decay channel.

\subsection{Tunneling in a cubic potential}

\begin{figure}[h]
\centering\includegraphics[width=0.9\textwidth]{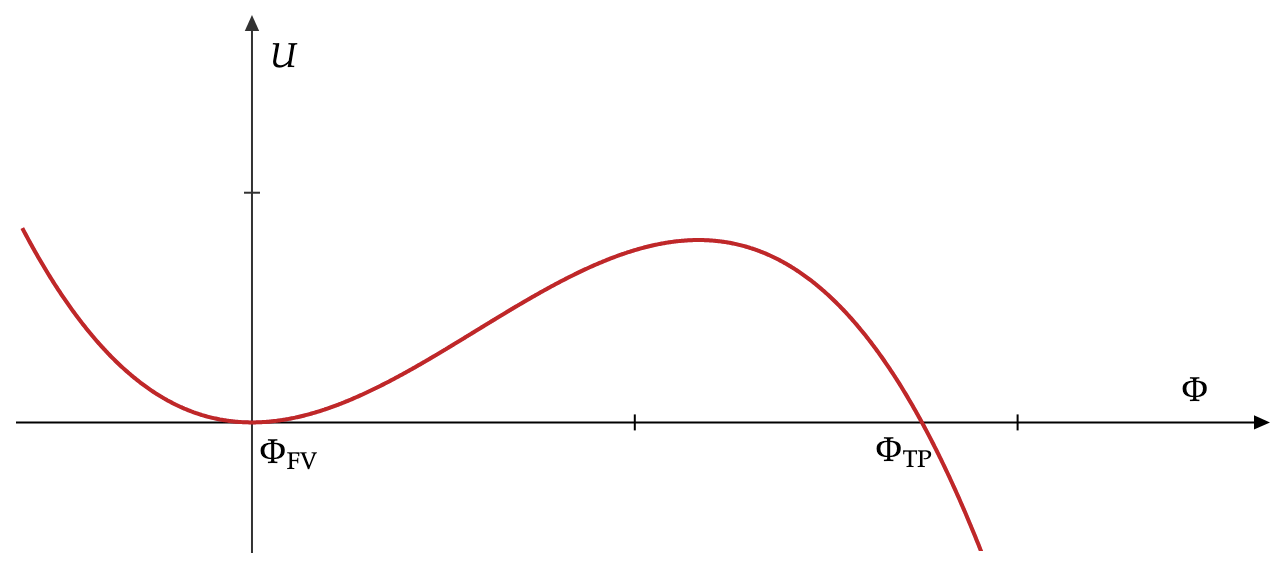}
\caption{Plot showing schematically the potential of eq. \eqref{eq:cubicpot}. The field can tunnel from the false vacuum at $\Phi=0$ to $\Phi_{\mathrm{TP}}=3\mu^2/2g$.}
\label{fig:cubicpot}
\end{figure}

While bounce solutions for general potentials are difficult to solve, a cubic potential leads to tunneling with a bounce profile that is analytically solvable, and we can obtain the exact lowest eigenvalues. So we will solve for such a potential below, and attempt to infer general lessons from our results. Consider the following Euclidean action for a scalar field:.
\begin{equation}
     I_\mathrm{E}[\Phi]=\int d\tau \ d^d x \ \Bigg(\frac{1}{2}(\partial_\tau\Phi)^2+\frac{1}{2}(\nabla_\mathbf{x}\Phi)^2 + U(\Phi)\Bigg),
\end{equation}
where (see figure \ref{fig:cubicpot})
\begin{equation}\label{eq:cubicpot}
    U(\Phi)=\frac{\mu^2}{2}\Phi^2-\frac{g}{3}\Phi^3.
\end{equation}

Even in infinte space, a field initially at $\Phi_{\mathrm{FV}}$ will tunnel to a value to the right of $\Phi_{\mathrm{TP}}$ via Coleman's $O(D)$ bounce, where $D=d+1$, and will then roll classically in real time. There will be a critical bubble of size $r_\star$. It is clear from eq. \eqref{eq:SchwingerRateCompact} that we expect the homogeneous solution to be valid so long as spatial sizes are smaller than $r_\star$. We will check this point explicitly below.

Before proceeding, let us take a quick detour on approximating the value of $r_{\star}$. The bounce equation of motion is
\begin{equation}
   \frac{d^2\Phi} {d\rho^2}+\frac{d}{\rho}\frac{d \Phi}{d\rho} =\mu^2 \Phi-g \Phi^3,
\end{equation}
and the solution must be of the form:
\begin{equation}
    \Phi(\rho)=\frac{\mu^2}{g} f(\mu \rho).
\end{equation}
We can see show this by direct substitution. Indeed, in these variables, the equation of motion is: 
\begin{equation}
   \frac{1}{\mu^2} \Bigg(\frac{d^2f} {d\rho^2}+\frac{d}{\rho}\frac{df}{d\rho} \Bigg)=f-f^2.
\end{equation}
Also, the boundary condition $f(\mu \rho)\rightarrow 0$ as $\rho\rightarrow \infty$ implies that for large $\mu \rho$,
\begin{equation}
    \frac{d^2f} {d\rho^2} \approx \mu^2 f,
\end{equation}
and so $f\sim\exp{(-\mu \rho)}$. Moreover, since for this (inverted) potential the particle must start rolling immediately, there is no thin-wall type configuration; the bounce is necessarily of thick-wall type.  It therefore follows that $r_\star \sim \mu^{-1} $, up to $\mathcal{O}(1)$ factors.

From now on, consider space to be compact and smaller than this critical bubble. Using the methods of section \ref{sec:compact}, we again seek homogeneous solutions to the energy conservation equation:
\begin{equation}
    \frac{1}{2}\dot{\Phi}^2=\frac{\mu^2}{2}\Phi^2-\frac{g}{3}\Phi^3,
\end{equation}
satisfying $\dot{\Phi}(0)=0$ and $\Phi(\tau)\rightarrow \Phi_{\mathrm{FV}}=0$ as $\tau\rightarrow \pm \infty$. Note though, that even without explicitly solving this equation, the roots of the potential are enough to find the decay rate using eq. \eqref{eq:compactrate}:
\begin{equation}
    \Gamma \sim \exp{\Bigg(-\frac{6}{5}\frac{\mu^5 V}{g^2}\Bigg)}.
\end{equation}

Nevertheless, the homogeneous bounce solution is:
\begin{equation}\label{eq:PTfield}
    \tilde{\Phi}(\tau)=\frac{3\mu^2}{2g}\frac{1}{\cosh^2{(\mu\tau/2)}}=\frac{\Phi_{\mathrm{TP}}}{\cosh^2{(\mu\tau/2)}}.
\end{equation}
Using the same time translation symmetry arguments as above, it is a simple matter to show that $\dot{\tilde{\Phi}}\sim \tanh(\mu\tau/2)/\cosh^2{(\mu\tau/2)}$ is an eigenfunction - of vanishing eigenvalue - of $\tilde{\mathcal{O}}$, which now has $U''(\tilde{\Phi})=\mu^2-2g\tilde{\Phi}$. Note that $\dot{\tilde{\Phi}}$ is manifestly antisymmetric in $\tau$ with only a single node, and so it must have the second lowest eigenvalue, which we have shown is zero. Again, this fact implies that there is exactly one negative eigenvalue corresponding to some eigenfunction with no nodes. A direct substitution shows that $\tilde{\Phi}^{3/2}$ is precisely that eigenfunction, with eigenvalue $-5\mu^2/4$. 

\begin{figure}[h]
\centering\includegraphics[width=0.85\textwidth]{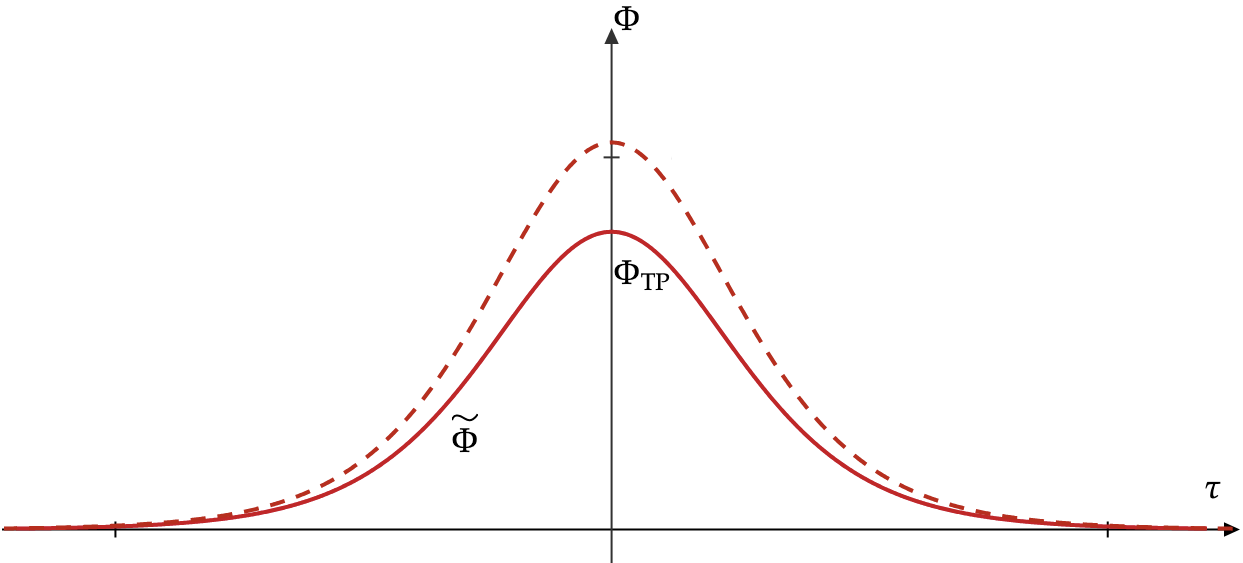}
\caption{A transformation of the bounce solution $\tilde{\Phi}(\tau)$ corresponding to the negative eigenvalue of $\tilde{\mathcal{O}}$. The solid red line shows the original homogeneous bounce solution $\tilde{\Phi}(\tau)$; the dashed red line shows the transformed field, i.e. with a perturbation proportional to the eigenfunction ($\tilde{\Phi}+\epsilon\tilde{\Phi}^{3/2}$) corresponding to the negative eigenvalue. }
\label{fig:dilatephi}
\end{figure}

At this point, it is worth considering the physical meaning of this negative mode. Consider the following perturbation:
\begin{equation}
    \tilde{\Phi}\rightarrow \tilde{\Phi}+\epsilon\tilde{\Phi}^{3/2}.
\end{equation}
We would like to compare this with Coleman's result that the negative eigenvalue corresponds to the transformation of dilating the $O(D)$ bubble \cite{Coleman:1977py, Coleman:1977th,Callan:1977pt}. Figure \ref{fig:dilatephi} suggests a physical interpretation: since the perturbation only marginally changes the value of the field far from the bounce center (owing to the exponential suppression: $\tilde{\Phi}(\tau)^{3/2}\sim\exp{(-3\mu |\tau|/2)}$ for large $|\tau|$), the only significant modification to the bounce profile is near its center $\tau=0$. But $\tilde{\Phi}(0)=\Phi_{\mathrm{TP}}$, and so this transformation is almost tantamount to just shifting the position of the turning point in figure \ref{fig:cubicpot}, but \textit{without} shifting the field profile much elsewhere. That is to say, the corresponding transformation associated with the negative eigenvalue amounts to changing the depth of barrier penetration in the tunneling process.

It is worth observing a curious mapping between the problem of finding negative eigenvalues of the quadratic fluctuation operator $\tilde{\mathcal{O}}$ and the problem of potential scattering in non-relativistic quantum mechanics in $1d$. To see this equivalence, the eigenvalue problem of the operator $\tilde{\mathcal{O}}$ amounts to finding solutions of
\begin{equation}
    \Big(-\partial_\tau^2 + U''(\tilde{\Phi}(\tau))\Big)\Psi(\tau)=\lambda\ \Psi(\tau).
\end{equation}

If we view the second derivative of the potential $U(\tilde{\Phi})$ as a potential in its own right and label it $W(\tau)=U''(\tilde{\Phi}(\tau))$, then this equation becomes

\begin{equation}
    \Big(-\partial_\tau^2 + W(\tau)\Big)\Psi(\tau)=\lambda\ \Psi(\tau).
\end{equation}

This expression is a Schrödinger equation of a particle scattering in an effective potential $W(\tau)$, if we view $\tau$ as a position variable. We should therefore interpret the fluctuations $\xi\rightarrow\Psi$ as wavefunctions for this particle, and $\lambda$ as its energy levels. As such, to show that the homogeneous solution $\tilde{\Phi}$ gives a valid decay channel, it is enough to check that the potential $W(\tau)$ evaluated on the bounce solution supports only a single negative energy state. In the example of eq. \eqref{eq:cubicpot}, the effective potential is
\begin{equation}\label{eq:EffectivePotential}
    W(\tau)=\mu^2 -\frac{3\mu^2}{\cosh^2(\mu\tau/2)}.
\end{equation}
Eq. \eqref{eq:EffectivePotential} is a special case of the Pöschl-Teller potential, and it has only one negative energy eigenvalue $\lambda=-5\mu^2/4$, and ground state wavefunction $\Psi(\tau)\sim \tilde\Phi(\tau)^{3/2}$.

\subsection{The full eigenvalue spectrum}

We now return to finding the eigenvalues of $\mathcal{O}$ in the case of eq. \eqref{eq:quadOperator}. We are especially interested in its lowest eigenvalues, which are related those of $\tilde{\mathcal{O}}$ which we have been looking at so far. First, let us label the eigenvalues of $\tilde{\mathcal{O}}$ by the list $\{\lambda_i\}=\{\lambda_{-},\lambda_0,\lambda_1,\lambda_2, \cdots \}$ in ascending order, with $\lambda_{-}$ being the negative eigenvalue, $\lambda_0=0$ being the vanishing eigenvalue, and $\lambda_1, \lambda_2, \cdots$ being the positive eigenvalues. For instance, for the cubic potential above (and eq. \eqref{eq:EffectivePotential}), the spectrum is:
\begin{equation}
    \lambda_-=-5\mu^2/4, \quad \lambda_0=0, \quad \lambda_1=3\mu^2/4, \quad \cdots,
\end{equation}

The Laplacian in $\mathcal{O}$ is sensitive to the number of spatial dimensions and the geometry of space. For example in $(1+1)d$, with space compactified on a circle of circumference $L$, we can decompose the field deviation of eq. \eqref{eq:fieldExpansion} in the Fourier basis:
\begin{equation}\label{eq:fourierexpansion}
    \xi(\tau,x)=\sum_n \xi_n(\tau)\exp{(i\ 2\pi n x/L)},
\end{equation}
with $\xi_n(\tau)$ being an eigenfunction of $\tilde{\mathcal{O}}$. Perturbing $\tilde{\Phi}$ by the $n$-th Fourier mode above, we see that \textit{each} eigenvalue of the list $\{\lambda_i\}$ increases to
\begin{equation}\label{eq:eigenvalueShift}
    \lambda_i \rightarrow \lambda_i+\Big(\frac{2\pi n}{L}\Big)^2.
\end{equation}
Since $n$ is any integer, the positive eigenvalues increase and hence remain positive for arbitrary $n$. The zero eigenvalue stays as such for $n=0$ but increases for any other $n$. The interesting change is in the negative eigenvalue $\lambda_- \rightarrow \lambda_- +(2\pi n/L)^2$. If space is compactified on a sufficiently small circle, then any non-zero $n$ will give a positive eigenvalue. In that case, $n=0$ corresponds to $\lambda_-$ still remaining the only negative eigenvalue, and the homogeneous solution remains a valid decay channel. A subtlety occurs when $L$ is sufficiently large. Then for low values of $n$, $\lambda_- +(2\pi n/L)^2$ may be negative, leading to an extremum of the action with multiple unstable directions rather than just one, and so we may not be able to regard this solution as a decay channel. Instead, we should find a new saddle point solution that has only one negative eigenvalue. 

As an explicit example, take the problem of the cubic potential in eq. \eqref{eq:cubicpot}. There we found that $\lambda_- = -5\mu^2/4$. Ensuring that this eigenvalue remains the only negative one on a spatial circle of circumference $L$ requires having $\mu L<4\pi /\sqrt{5}$. \footnote{Note that we do \textit{not} require that $\mu L \ll 4\pi /\sqrt{5}$. Even barely satisfying the inequality is enough to ensure a single negative eigenvalue and thus an bounce that mediates a decay channel.}

There is another point to mention regarding the lowest eigenvalues. If the spatial circle has circumference $\mu L=4\pi /\sqrt{5}$, then there are now two vanishing eigenvalues: the original $\lambda_0$, and the new $\lambda_-+(\pm 2\pi/L)^2$. The second is doubly degenerate; one combination of their eigenfunctions is a collective coordinate which stays 0 for larger $L$ whereas the other indicates a start of an instability to an inhomogeneous solution. Since zero modes are associated with collective coordinates related to translation in Euclidean spacetime, it seems that this extra mode indicates that we must include a correction to the homogeneous solution $\tilde\Phi$ involving a higher Fourier mode in eq. \eqref{eq:fourierexpansion}. This addition breaks homogeneity in space and there must be a collective coordinate associated with translating the center of mass of the new solution along the $x$ direction.

Moreover, we could make a similar argument in $(2+1)d$ where space is a sphere of radius $R$. Then it is natural to decompose the deviation using spherical harmonics $Y_{lm}$:
\begin{equation}
    \xi(\tau,\theta,\phi)=\sum_{l,m}\xi_{lm}(\tau)Y_{lm}(\theta,\phi).
\end{equation}
Just as in eq. \eqref{eq:eigenvalueShift}, the tower of eigenvalues is now shifted once more:
\begin{equation}
    \lambda_i \rightarrow \lambda_i+\frac{l(l+1)}{R^2}.
\end{equation}
For the problem of tunneling through a cubic potential barrier (eq. \eqref{eq:cubicpot}), demanding that there be no more than a single negative eigenvalue requires $\mu R<\sqrt{8/5})$. Now in the limiting case where $\mu R=\sqrt{8/5}$, the $l=1$ mode corresponds to a vanishing eigenvalue. However, for each $l$, there are $(2l+1)$ degenerate eigenfunctions, so there must now be two new collective coordinates associated with the rotating the center of mass - of an inhomogeneity from a higher harmonic - on the sphere. The other one indicates a start to an instability of an inhomogeneous correction, for ever so slightly larger $R$.

\subsection{Real time fluctuations}

It is worth spending a moment on trying to understand the calculations in the previous sections in terms of what is happening in real time, at least qualitatively. There has been much recent work on understanding false vacuum decay in real time \cite{Braden:2018tky, Batini:2023zpi}. In particular, \cite{Braden:2018tky} demonstrates that Coleman's calculation of the Euclidean bounce actually computes the likelihood of quantum fluctuations on a false vacuum background forming a coherent state of a bubble with a true vacuum interior. The exponential suppression of the large bounce action captures the extremely low probability of quantum fluctuations adding up to a coherent state. We would like to understand what the bounce action suppression corresponds to in the case of small finite volume since we now no longer have a coherent state of bubble to produce. 

Consider once again the potential of eq. \eqref{eq:cubicpot} with the potential shown in figure \ref{fig:cubicpot}. Take space to be a 2-sphere of radius $R$. In the distant past, the scalar field has expectation value $\langle \Phi \rangle=0$. Near the false vacuum, we may use a quadratic approximation (with $\Phi=\langle \Phi \rangle+\delta\Phi$):
\begin{equation}
    I[\delta\Phi]\approx \int dt\ d^2 x\ \Bigg(-\frac{1}{2}(\partial_\mu \delta\Phi)^2-\frac{\mu^2}{2}(\delta\Phi)^2\Bigg).
\end{equation}
Defining 
\begin{equation}
    \delta\Phi(t,\theta,\phi)=\sum_{l,m}Q_{lm}(t)Y_{lm}(\theta,\phi),
\end{equation}
the action decomposes into an infinite sum over harmonic oscillators with a hierarchy of frequencies:
\begin{equation}
    I[\{Q_{lm}\}]\approx \frac{R^2}{2}\int dt\ \sum_{l,m}\Bigg( |\dot{Q}_{lm}|^2-\mu_l^2|Q_{lm}|^2 \Bigg),
\end{equation}
where 
\begin{equation}
    \mu_l^2=\mu^2+\frac{l(l+1)}{R^2}.
\end{equation}
Now, upon quantizing by imposing the usual canonical commutation relations for harmonic oscillators, we can compute the fluctuations of the the $Q_{lm}$'s in the ground state:
\begin{equation}
    \langle Q_{lm}^2 \rangle =\frac{1}{2R^2 \mu_l} = \frac{1}{2R\sqrt{(\mu R)^2+l(l+1)}},
\end{equation}
while $\langle Q_{lm}\rangle=0$. 

Now suppose $\mu R \lesssim 1$. Then the $l=0$ mode fluctuates substantially ($\langle Q_{00}^2\rangle\sim 1/\mu R^{2}$), whereas the higher $l$ modes fluctuate less ($\langle Q_{lm}^2\rangle\sim 1/(R\sqrt{l(l+1)})$) and are comparatively frozen out, as $R$ decreases. As such, the homogeneous mode alone is likely to tunnel out of the false vacuum value to the true vacuum value by rapid fluctuation. In imaginary time, the Eucildean homogeneous bounce action computes the chances of this fluctuation causing a tunnelling event. Note that the end state of such a fluctuation is a homogeneous configuration with the field now in the true vacuum, \textit{since it is only the homogeneous field that tunnelled out} as its fluctuations were largest. If we make $R$ larger and larger, higher $l$ modes begin to fluctuate rapidly so long as $l(l+1)\lesssim (\mu R)^2$, but even higher $l$ modes remain frozen out. This point clarifies the above discussion - based on analyzing the eigenvalue spectrum - on why inhomogeneous corrections are necessary when we increase the size of space. 

Note also that a similar qualitative argument follows in the case of space being circle of circumference $L$. Decomposing the field in the Fourier basis,
\begin{equation}
    \Phi(t,x)=\sum_n Q_n(t)\exp{( i\ 2\pi n x/L)},
\end{equation}
with $\langle Q_{n}\rangle=0$, we once again obtain an action for an infinite sequence of harmonic oscillators, similar to the earlier discussion. We see that when $L \lesssim \mu^{-1}$, $\langle Q_{0}^2\rangle \sim 1/\mu L$, whereas $\langle Q_{n}^2\rangle \sim 1/|n|$ when $n$ is non-vanishing. Once again, at small physical sizes, only the homogeneous mode fluctuates rapidly while the higher modes are effectively frozen out.

One final point to make is that once the decay occurs, the field then takes the initial value conditions $\Phi(0,\mathbf{x})=\Phi_{\mathrm{TP}}$ and $\dot{\Phi}(0,\mathbf{x})=0$, which one must use to evolve the real time equations of motion; see figure \ref{fig:CompactInstanton}. In doing so, the effective frequencies of the quadratic fluctuations $\xi$ in eq. \eqref{eq:actionQuadratic} will change in time, leading to the formation of quanta associated with these fluctuations. In essence, the physical picture is the following: a scalar field tunnels so that a false vacuum decays homogeneously to a true vacuum of lower energy, producing quanta of excitations on top of the true vacuum background. This picture is strikingly similar to the Schwinger effect on small compact space \cite{brown2018schwinger}; see the discussion after eq. \eqref{eq:SchwingerRateCompact2}.

\section{Revisiting the Schwinger effect in compact space}
\label{sec:appA}

We are going to carry out a non-trivial test of our compact space decay rate formula in eq. \eqref{eq:SchwingerRateCompact2}. In particular, a theory mimicking axion electrodynamics in $(1+1)d$ presents a solvable model in which we may verify eq. \eqref{eq:SchwingerRateCompact2}. We will be brief here; see \cite{Hassan:2024nbl, Hassan:2024zkd} for a more detailed review of this model.

The Lorentzian signature action for $(1+1)d$ axion electrodynamics is 
\begin{equation}
    \label{eq:amaction2d}
    I=\int d^2 x \Bigg( -\frac{1}{4}F_{\mu \nu}F^{\mu \nu} - \frac{1}{2}\partial_{\mu}\theta \partial^{\mu}\theta -m^2 \Big(1-\cos \theta\Big) +  \frac{e}{2\pi} \theta \epsilon_{\mu \nu } F^{\mu \nu}\Bigg).
\end{equation}
Here, $F_{\mu\nu}$ is the Maxwell field strength for the photon $A_\mu$, and the pseudoscalar $\theta$ is the axion field. Also, $e$ is the electromagnetic coupling. The antisymmetric symbol is $\epsilon_{\mu\nu}$, with $\epsilon_{01}=+1$. 
$\theta$ is periodic with the gauge identification $\theta \sim \theta + 2\pi$. 
In (1+1)$d$, there is only one independent component of the Maxwell tensor which is the electric field $E = F_{10}$. Varying the action eq. \eqref{eq:amaction2d} with respect to the fields gives:
\begin{align}
\label{eq:eom2d}
\partial_x E =-\frac{e}{\pi}\partial_x \theta, \quad \partial_t E =-\frac{e}{\pi}\partial_t \theta,  \mathrm{\ and,} \\
\frac{1}{m^2}\partial^2 \theta =\sin\theta-\frac{eE}{\pi m^2}.\label{eq:eom2d2}
\end{align}

From eq. \eqref{eq:eom2d2} constant $\theta$ local extrema are given by 
\begin{equation}
\theta=\arcsin{\Bigg(\frac{eE_0}{\pi m^2}\Bigg)}, 
\end{equation}
where $eE_0/\pi m^2 \leq 1$, with subsequent minima $2\pi$ units apart. Let us now specialize to the case of extremely weak coupling in a constant background electric field, or more precisely, consider $e\rightarrow 0, E_0 \rightarrow\infty$ while keeping $eE_0$ fixed; additionally, we take $eE_0/m^2\ll 1$, which is stronger than the earlier requirement. Call the first minimum $\theta_0\approx eE_0/\pi m^2$. Note that the energy difference between the minima is now much lower than the height of the potential barrier, and so we anticipate soon seeing thin wall solutions. We now discuss tunneling between these minima. In Euclidean signature, the action is
\begin{align}\label{eq:euclideanaction2d}
    I_{\mathrm{E}}=\int dx\  d\tau \Bigg( \frac{1}{4}F_{\mu \nu}^{(\mathrm{E})}F_{\mu \nu}^{(\mathrm{E})} +\frac{1}{2}(\partial_{\mu} \theta)^2 +m^2 \Big(1-\cos \theta\Big)
    -i\frac{e}{2\pi}\theta\epsilon^{\mu \nu}F_{\mu \nu}^{(\mathrm{E})} \Bigg),
\end{align}
where the superscript denotes Euclidean fields $F^{(\mathrm{E})}_{\mu\nu}=\pd_\mu A^{(\mathrm{E})}_\nu - \pd_\nu A^{(\mathrm{E})}_\mu$, with $iA^{(\mathrm{E})}_{\tau}=A_{0}$; also note that $\epsilon_{1 \tau}=1$ whereas previously in Lorentzian signature $\epsilon_{01}=1$. Now, if the electric field is large, then we can neglect its changes which are proportional to $e$. Then we can search for a solution to the equation of motion for $\theta$, and one exists which has $O(2)$ symmetry\footnote{Here, $\rho^2/m^2=\tau^2+x^2$.}:
\begin{equation}\label{eq:O2thinwall}
    \theta(\rho)=\theta_0+2\pi-4\arctan{\Big(\exp(\rho-\rho_\star)\Big)}.
\end{equation}

The wall thickness is therefore $\mathcal{O}(m^{-1})$, and the walls are situated at $x=\pm m^{-1} \rho_\star$ upon nucleation. We may find $\rho_\star$ by substituting this solution in eq. \eqref{eq:euclideanaction2d}, and minimizing to obtain the radius of the critical bubble: 
\begin{align}
\rho_\star \approx \frac{4m^2}{eE_0}. 
\end{align}

An important point is that the domain walls carry electric charge according to eq. \eqref{eq:eom2d}, with $Q_{\mathrm{DW}}=\pm 2e$. One can also calculate the mass of the domain walls from the energy momentum tensor, obtainable from eq. \eqref{eq:amaction2d}; we obtain $m_{\mathrm{DW}}=8m$. Putting eq. \eqref{eq:O2thinwall} in the Euclidean action eq. \eqref{eq:euclideanaction2d}, we find the bounce action $I_{\mathrm{B}}= 32\pi m^2/e E_0$. In domain wall variables, this expression yields
\begin{equation}
    \Gamma/L\sim \exp\left(-\frac{\pi m_\mathrm{DW}^2}{Q_\mathrm{DW}E_0}\right).
\end{equation}
The Schwinger exponent is manifest here. To summarize, the bubble walls are domain walls carrying electric charge and have fixed mass. These are kinks in $1d$ and hence mimic particles. They spontaneously nucleate from an electric field background. This process is therefore a type of Schwinger effect. This outcome is largely expected from earlier work on bosonization of the Schwinger model in $(1+1)d$ (see also \cite{Coleman:1974bu,Coleman:1985rnk,Coleman:1976uz,Mandelstam:1975hb,Schwinger:1962tp}).

Now we would like to see what happens when we compactify the $x$ direction on a circle of circumference $L \ll \rho_\star$. Realizing that this process is a Schwinger effect, intuition suggests - based on eq. \eqref{eq:SchwingerRateCompact2} - the following rate in compact space:
\begin{equation}\label{eq:dwrate}
    \Gamma \sim \exp{\Big(-2m_{\mathrm{DW}}L \Big)},
\end{equation}
where we have simply replaced the parameter $m$ with $m_{\mathrm{DW}}$ since we are viewing the domain walls as particles of definite mass.

The special case of eq. \eqref{eq:compactrate} in $(1+1)d$ is 
\begin{equation}\label{eq: 1dcompactrate}
    \Gamma \sim \exp{\Bigg( -2L\int_{\Phi_{\mathrm{TP}}}^{\Phi_{\mathrm{FV}}}d\tilde{\Phi}\sqrt{2U(\tilde{\Phi})} \Bigg)}.
\end{equation}
We must now test, using eq. \eqref{eq: 1dcompactrate}, to see if the prediction of eq. \eqref{eq:dwrate} matches. A direct application of eq. \eqref{eq: 1dcompactrate}, using the potential in eq. \eqref{eq:euclideanaction2d}\footnote{A moment's thought shows that since $eE_0$ is held fixed and $\theta_0\ll 1$, the last term in eq. \eqref{eq:euclideanaction2d} will give at most an $\mathcal{O}(\theta_0)$ correction to the potential. As usual, the action in the absence of a bubble is normalized to 0 which must be taken into account due to the background electric field.}, gives the approximate\footnote{We use the fact that $\theta_0\ll 1$ and so the first and second minima are almost at $0$ and $2\pi$ respectively.} exponent
\begin{equation}
    2L\int_{0}^{2\pi}\sqrt{2m^2(1-\cos\theta)}=2\times 8m\times L=2m_{\mathrm{DW}}L,
\end{equation}
with corrections of $\mathcal{O}(\theta_0)\ll 1$. Hearteningly, eq. \eqref{eq: 1dcompactrate} agrees with what we expect from eq. \eqref{eq:SchwingerRateCompact2} and eq. \eqref{eq:dwrate}, and this completes our check.

\section{Conclusion and outlook}
\label{sec:conclusion}

We have explored the effects of finite size on false vacuum decay processes. When space is compactified on scales smaller than the size of Coleman's usual critical bubble, a quantum mechanical bounce mediates the decay process from false vacuum to true vacuum. This time, the qualitative behavior is different from the usual process of bubble nucleation. Indeed, since no critical bubble fits in space, the entire field decays homogeneously out of the false vacuum configuration. Nevertheless, intricate features such as the presence of vanishing modes and a negative mode in the quadratic fluctuation operator show similarities between the quantum mechanical Euclidean bounce and the usual Euclidean bounce in field theory. We have shown this with an exactly solvable cubic potential problem, including by checking the eigenvalue spectrum. Moreover, we have also connected this this process to earlier worldline treatments of the Schwinger effect in compact space and have shown that homogeneous bounces give the expected decay rate for the Scwhinger effect in finite volume.

There are several interesting directions for exploration, including in condensed matter systems and model-building in early universe cosmology, as well as extensions of this present work. For example, it would be interesting to see the effects of higher inhomogeneous modes and their correction effects to homogeneous tunneling; such a system is likely to mimic WKB tunneling of a particle in a multi-dimensional potential \cite{Bender:1969si,Bender:1973rz}. This topic also lends itself to real time simulations, among other interesting avenues of research.

\section*{Acknowledgements}
SH is a Junior Research Fellow of Christ Church, University of Oxford, and is grateful for their support. We thank Georges Obied for discussions of the Schwinger effect and earlier collaboration on axion electrodynamics.  For the purpose of Open Access, the author has applied a CC BY public copyright license to any Author Accepted Manuscript version arising from this submission.

\appendix

\section{Review of the Schwinger effect in compact space}\label{sec:ScwhingerCompactSpace}
We summarize here some key results on the Scwhinger effect in compact space. We will briefly outline this process in the worldline formalism following \cite{brown2018schwinger, Ai:2020vhx}, which allows for immediate generalization to compact space. 

\begin{figure}[h]
\centering\includegraphics[width=0.85\textwidth]{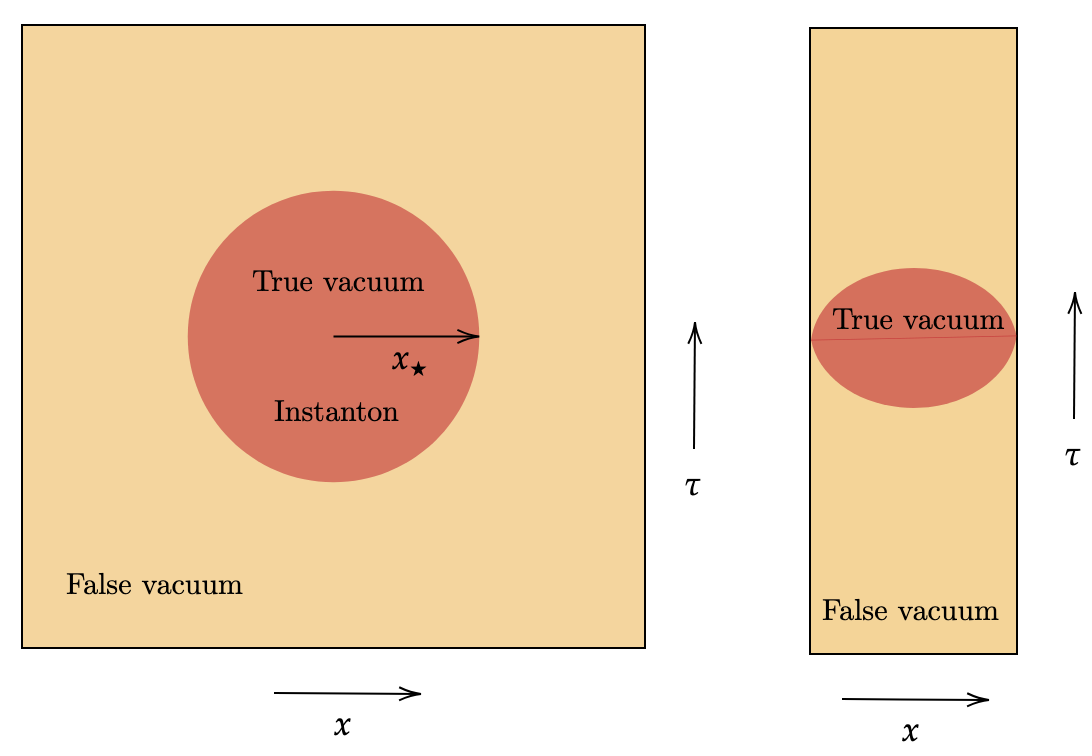}
\caption{Left: The $O(2)$ bounce solution for the Schwinger effect. The red bubble represents the true vacuum solution (a critical bubble of radius $x_\star$ containing a lower electric field) than the false vacuum exterior shown in beige. Right: The Schwinger effect in compact space (smaller than the critical bubble) showing a lens shaped bounce.}
\label{fig:figSchwinger}
\end{figure}

In the usual Schwinger process (we focus on $(1+1)d$ here for simplicity), a background electric field decays via pair production of an electron and positron, oriented such that the intermediate electric field between them is screened. The field then accelerates these pairs outwards, and the intermediate lower electric field expands. It is natural to view this process as one of false vacuum decay, with the particles akin to bubble walls, the interior resembling a true vacuum, and the exterior a false vacuum. To see this quantitatively, consider the Euclidean worldline (WL) action for a pair of mass $m$ and charge $\pm q$ in a background electric field $E$:
\begin{equation}
    \label{eq:EuclideanWL}
    I_\mathrm{E}=m\int_{\mathrm{WL}} d\tau \  (1+\dot{x}(\tau)^2)^{1/2}-qE\int_{\mathrm{WL}} d\tau \ x(\tau),
\end{equation}
where $\tau$ is the Euclidean time (\textit{not} the proper time), $x(\tau)$ is half the relative separation of the pair, and an overdot indicates differentiation with respect to $\tau$. The first integral involves an evaluation of a perimeter and the second an evaluation of an area\footnote{See \cite{brown2018schwinger} for additional details.}. Varying the action with respect to $x(\tau)$ gives the equation of motion, whose solution is $O(2)$ symmetric:
\begin{equation}
    \label{O2solution}
    x^2+\tau^2=x_{\star}^2,
\end{equation}
where $x_\star=m/qE$ is the critical radius of the nucleated bubble. See the left panel of figure \ref{fig:figSchwinger}. Putting this solution in eq. \eqref{eq:EuclideanWL}, we obtain the decay rate per unit length (eq. \eqref{eq:SchwingerRate}):
\begin{equation}
    \Gamma/L \sim \exp{\Bigg(-\frac{\pi m^2}{qE}\Bigg)}.
\end{equation}
\\

Moreover, upon Wick rotating to real time ($\tau=it$), eq. \eqref{O2solution} becomes:
\begin{equation}
    x^2-t^2=x_\star^2,
\end{equation}
which represents a hyperbola in $(t,x)$ coordinates. Physically, this expression corresponds to the particle pair accelerating outwards due to the background electric field.

Now consider the case in which space is compactified on a circle. Moreover, take the perimeter of this circle, $L$, to be less than the size of the critical bubble. It is then clear that the $O(2)$ critical bubble bounce cannot be accommodated. Nevertheless, eq. \eqref{O2solution} continues to be a solution\footnote{With the exception of the endpoint cusps.} to the equation of motion and hence a saddle point of the action. The solutions are therefore geometric segments of circle of radius $x_\star$ patched together\footnote{This solution is perhaps unsurprising since in Euclidean spacetime, swapping compact size for compact time corresponds to false vacuum decay at finite temperature, and this lens shaped bounce appears in such cases \cite{Garriga:1994ut}.}. See the right panel of figure \ref{fig:figSchwinger}. Owing to the fact that evaluating the bounce action merely requires a computation of perimeter and area (eq. \eqref{eq:EuclideanWL}), and that the right panel of figure \ref{fig:figSchwinger} shows two segments joined together, we can find the decay rate:
\begin{equation}
\label{eq:SchwingerRateCompact}
\begin{split}
    \Gamma & \sim \exp{\Bigg(- \frac{2m^2}{qE}\arcsin{\Big(\frac{L}{2x_\star}\Big)-mL\Big(1-\frac{L^2}{4x_{\star}^2}\Big)^{1/2}\Bigg)}} \\
    & \approx \exp{\Bigg(-2mL\Big( 1-\frac{ L^2}{24 x_\star^2} +\cdots\Big)\Bigg)},
\end{split}
\end{equation}
where the last step holds when we seek only the leading effect of finite size.

\section{Finite size enhancement of decay rates on a sphere}
\label{sec:appB}

To see how compact space effects can accelerate bubble nucleation rates, it is especially insightful to study the problem in $(2+1)d$, with space being a $2-$sphere. 

Before doing so, we first outline some key points regarding the decay of the false vacuum on an infinite plane. 

\subsection{On the infinite plane}

Consider the action of a real scalar field $\Phi$ on an infinite plane:
\begin{equation}
    I=\int d^3 x\ \Big(-\frac{1}{2}\partial_\mu \Phi \partial^\mu \Phi -U(\Phi)\Big),
\end{equation}
with
\begin{equation}
    U(\Phi)=U_{+}(\Phi)+\frac{\epsilon}{2a}(\Phi-a), \quad U_{+}(\Phi)=\frac{\lambda}{8}\Big(\Phi^2 -\frac{\mu^2}{\lambda}\Big)^2,
\end{equation}
and $\mu^2/\lambda=a^2$. We take $\epsilon\ll \mu^4/\lambda$, so that the potential barrier of this quartic potential far exceeds the difference in energy levels of the two minima (see figure \ref{fig:potential}; note that in this case $\Phi_{\mathrm{FV}}$ and $\Phi_{\mathrm{TV}}$ are at approximately $+a$ and $-a$ respectively). 

Now, on an infinite plane, there is a critical bubble solution (a string) that is $O(3)$ symmetric, whose radius is $r_\star=2\mu^3/3\lambda\epsilon$, with corresponding bounce action $I_{\mathrm{B}}=16\pi \mu^9/81 \epsilon^2 \lambda^3$. For what follows, it is convenient to express quantities in terms of the string tension $T=\mu^3/3\lambda$, which comes from computing the energy momentum tensor and evaluating it on the string solution. Then the critical radius is $r_\star=2T/\epsilon$, and the bounce action is $I_{\mathrm{B}}=16\pi T^3/3\epsilon^2$.

We will now set up the basic ingredients for generalizing the above results to quantum fields on a curved space. We can do so by recognizing that the effective degrees of freedom involve the coordinates of the nucleated string, as well as the energy gap between the interior and exterior. As such we can write a worldsheet effective theory (for instance, see \cite{Garriga:1991tb,Garriga:1991ts,Garriga:1992nm,Widrow:1989fe,Hassan:2024pht,Feng:2000if}) akin to a Nambu-Goto action:
\begin{equation}\label{eq:NGaction}
    I_{\mathrm{E}}=T \int_{\mathrm{WS}}d^2 \xi \sqrt{h}-\epsilon\int d\tau\ \int_{\mathrm{Interior}}d^2 x\ \sqrt{g}.
\end{equation}
Here, $\xi$ and $x$ are coordinates intrinsic to the string and the bulk respectively\footnote{The line element is $ds^2=d\tau^2+dr^2+r^2 d\phi^2.$}. The first term represents the standard Nambu-Goto action with a worldsheet integral, and the second represents a bulk integral for the region enclosed in space by the string\footnote{Strictly speaking, we should view $T$ and $\epsilon$ as Wilson coefficients, but the fact that they coincide with the tension and energy density respectively of the problem at hand follows from matching, say, a thin wall solution's bubble size and nucleation rate.}. We should now regard eq. \eqref{eq:NGaction} as a generalization of the worldline action of eq. \eqref{eq:EuclideanWL}. Indeed, this similarity is more transparent when we write:

\begin{equation}
    [h_{ab}]=
    \begin{pmatrix}
1+\dot{r}(\tau)^2 & 0 \\
0 & r(\tau)^2 
\end{pmatrix},
\end{equation}
so the action eq. \eqref{eq:NGaction} becomes
\begin{equation}
    I_{\mathrm{E}}\rightarrow  T \int d\tau\ d\phi\ r(1+\dot{r}^2)^{1/2}-\epsilon\int d\tau \int d\phi\ dr\ r.
\end{equation}
This follows from projecting the indices of the bulk metric:
\begin{equation}
    h_{ab}=g_{\alpha\beta}\frac{\partial x^\alpha}{\partial \xi^a}\frac{\partial x^\beta}{\partial \xi^b},
\end{equation}
where the coordinates are $x^\alpha=\{\tau, r, \phi \}$, and $\xi^a=\{\tau, \phi \}$. The fact that $r=r(\tau)$ follows from the symmetry of the ansatz. In any case, the equation of motion has a solution of the form $r^2+\tau^2=r_\star^2$, with $r_\star=2T/\epsilon$. An important issue to note is that this solution for $r(\tau)$ is compatible with the following boundary conditions:
\begin{itemize}
    \item $\dot{r}(0)=0$, which follows from seeking a solution for a single bounce which has time reversal symmetry at the center of the bounce. In other words, this statement means that the critical bubble is nucleated with a vanishing initial speed;
    \item $r(0)=r_\star=2T/\epsilon$, which actually follows independently from energy conservation - by matching the string mass with the energy in the interior region - assuming the critical bubble is born at rest in accordance with the above requirement.
\end{itemize}

When we generalize the above calculation from an infinite plane to a sphere, we will no longer retain the $O(3)$ symmetry, but we will still be able to use time reversal symmetry and energy conservation to obtain the boundary conditions necessary to solve the equation of motion.

\subsection{On the sphere}

\begin{figure}[h]
\centering\includegraphics[width=0.4\textwidth]{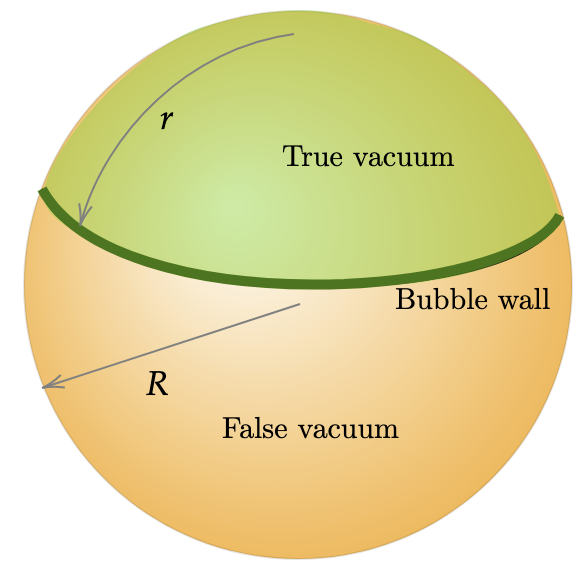}
\caption{A cap of true vacuum (green) nucleating on a spherical surface of false vacuum (beige), with a domain wall/string (dark green) delineating the boundary between the two regions.}
\label{fig:S2}
\end{figure}

We now calculate the bubble nucleation rate on a sphere. The great advantage of eq. \eqref{eq:NGaction} is that its generalization to curved space is immediate. For a sphere of radius $R$ (taking $R\gtrsim r_\star$), it is useful to define a \textit{radius} of a cap $r$ (see figure \ref{fig:S2}), with coordinate range $0\leq r \leq \pi R$.

In these coordinates, the line element is now
\begin{equation}
    ds^2=d\tau^2+dr^2+R^2\sin^2\Big(\frac{r}{R}\Big)d\phi^2.
\end{equation}
Along the lines of the earlier discussion on nucleating a string with a true vacuum interior on the flat plane, we now have a string of true vacuum interior nucleating as a cap on a sphere (see figure \ref{fig:S2}).

The action, using eq. \eqref{eq:NGaction}, is now:
\begin{equation}
    I_{\mathrm{E}}\rightarrow 2\pi R T \int d\tau\ (1+\dot{r}^2)^{1/2}\sin\Big(\frac{r}{R}\Big)- 2\pi R^2 \epsilon \int d\tau\ \Bigg(1-\cos\Big(\frac{r}{R}\Big)\Bigg).
\end{equation}
\begin{figure}[h]
\centering\includegraphics[width=0.7\textwidth]{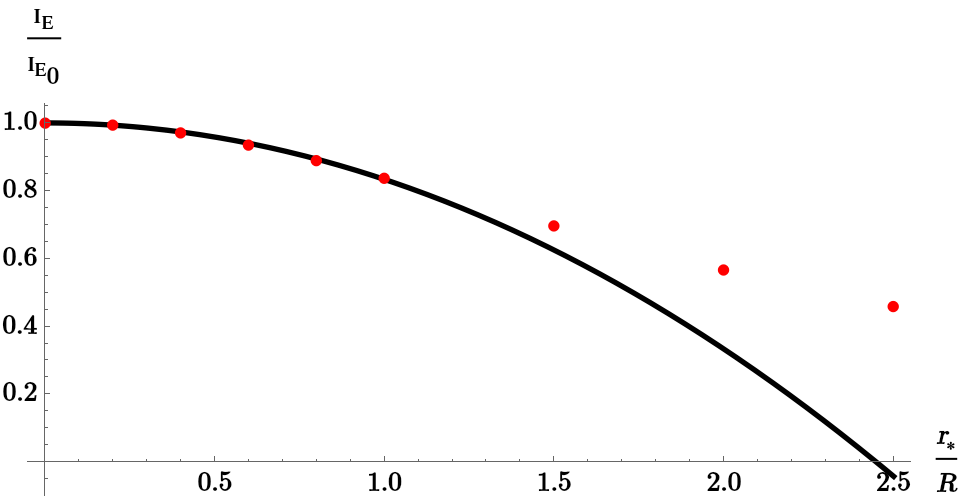}
\caption{The black line is given by the formula of eq. \eqref{eq:actionratioS2} (see text), and the red dots denote numerical results from integrating eq. \eqref{eq:EOMsphere}. The horizontal axis shows the ratio of a critical bubble's radius on a flat plane to the radius of the sphere, and the vertical axis gives the ratio of the Eucidean action (evaluated on the bounce solutions) on a sphere to that on a flat plane. }
\label{fig:instantononsphere}
\end{figure}

The equation of motion for $r(\tau)$ is:
\begin{equation}\label{eq:EOMsphere}
    R\frac{d}{d\tau}\Bigg(\frac{\dot{r}\sin \Big(\frac{r}{R}\Big)}{(1+\dot{r}^2)^{1/2}} \Bigg)-(1+\dot{r}^2)^{1/2}\cos \Big(\frac{r}{R}\Big)+\frac{R\epsilon}{T}\sin \Big(\frac{r}{R}\Big)=0.
\end{equation}
This time, a solution of the form $r^2+\tau^2=\mathrm{const.}$ will not work. We will have to numerically integrate eq. \eqref{eq:EOMsphere}, subject to the boundary conditions below:
\begin{itemize}
    \item $\dot{r}(0)=0$;
    \item $r(0)=2R\arctan(r_\star/2R)$. Note that $r_\star=2T/\epsilon$ is again the critical bubble size on the infinite plane (\textit{not} on the sphere). This conditions comes from energy conservation, i.e. equating the string mass with the energy difference in the cap interior (figure \ref{fig:S2}).
\end{itemize}

Before integrating eq. \eqref{eq:EOMsphere}, recall that there is formula for the bounce action on curved space ($I_{\mathrm{B}}$) relative to the bounce action on flat space ($I_{\mathrm{B}_0}$) that we had mentioned in section \ref{sec:compactchanges}. In particular, since the Ricci scalar for a sphere of radius $R$ is $2/R^2$, we can use eq. \eqref{eq:curvaturemodification} to get:
\begin{equation}\label{eq:actionratioS2}
    \frac{I_\mathrm{B}}{I_\mathrm{B_0}}\approx 1-\frac{r_\star^2}{6R^2},
\end{equation}
valid for $r_\star \lesssim R$. Figure \ref{fig:instantononsphere} depicts this ratio alongside numerically integrated values from eq. \eqref{eq:EOMsphere}. On a large sphere (small $r_\star/R$) this ratio of actions is near unity as expected, but the numerical results show deviations from eq. \eqref{eq:actionratioS2} when the radius of the sphere is close to $r_\star$. Indeed, numerical results rescue us from the nonsensical negative values of this ratio of actions  (for larger values of $r_\star/R$, at which point eq. \eqref{eq:actionratioS2} breaks down). As anticipated, the compactness of space appears to \textit{enhance} the nucleation rate by \textit{lowering} the bounce action.

\bibliographystyle{JHEP}
\bibliography{refs.bib}

\end{document}